\documentclass[twocolumn,showpacs,preprintnumbers,amsmath,amssymb,a4paper]{revtex4}

\usepackage{graphicx}% Include figure files
\usepackage{dcolumn}% Align table columns on decimal point
\usepackage{bm}% bold math

\usepackage{epsfig}
\usepackage{pifont}
\usepackage{wasysym}

\begin{document}
\preprint{}
%\draft

\title{Thermodynamics of DNA loops with long-range correlated structural disorder}
\author{C. Vaillant$^{1,2}$, B. Audit$^{3}$ and A. Arn\'eodo$^{3}$}
\address{$^{1}$Institut Bernouilli, EPFL, 1015 Lausanne, Switzerland\\
$^{2}$Laboratoire Statistique et G\'enome, 523 place des terrasses de l'Agora, 91000 Evry, France,\\
$^{3}$Laboratoire Joliot Curie, Ecole Normale Sup\'erieure de  Lyon, 46 all\'ee d'Italie, 69364 Lyon cedex 07, 
France.}      
\date{\today}

\begin{abstract}
We study the influence of a structural disorder on the thermodynamical 
properties of 2D-elastic chains submitted to mechanical/topological 
constraint as loops. The disorder is introduced via a spontaneous curvature 
whose distribution along the chain presents either no correlation or 
long-range correlations (LRC). The equilibrium properties of the one-loop 
system are derived numerically and analytically for weak disorder. 
LRC are shown to favor the formation of small loop, larger the 
LRC, smaller the loop size. We use the mean first passage time 
formalism to show that the typical short time loop dynamics 
is superdiffusive in the presence of LRC. Potential biological implications on 
nucleosome positioning and dynamics in eukaryotic 
chromatin are discussed.
\end{abstract}


\maketitle

The dynamics of folding and unfolding of DNA within living cells is 
of fundamental importance in a host of biological processes 
ranging from DNA replication to gene regulation~\cite{aVanHo88}. As the basic unit of 
eukaryotic chromatin organization, the structure and dynamics of nucleosomes 
has attracted increasing experimental and theoretical interest~\cite{aWidom99}. High 
resolution X-ray analyses \cite{aLuger97} have provided deep insight into the wrapping 
of $145~bp$ of DNA in almost two turns around an 
histone octamer to form a nucleosome core. Recent experiments have shown 
that nucleosomes are highly dynamical structures that can be moved along 
DNA by chromatin remodeling complexes~\cite{aPeter00} but that can 
also move autonomously on short DNA segments~\cite{aPenni91}.~Different models have been 
proposed to account for the nucleosome mobility~\cite{aSchie03} including the DNA 
reptation model that involves intranucleosomal loop diffusion~\cite{aSchie01} and the 
nucleosome repositioning model via an extranucleosomal loop~\cite{aMoham04}; both 
models provide an attractive picture of how a transcribing RNA polymerase can 
get around nucleosomes without dissociating it completely. Since the 
discovery of naturally  curved DNA~\cite{aMarin83}, several works have investigated 
the possibility that the DNA sequence may facilitate the nucleosome 
packaging~\cite{aIoshi96} in the same manner as it can highly promote very small loop
formation~\cite{aClout04}. Recently, a comparative statistical analysis of 
eukaryotic sequences and their corresponding DNA bending 
profiles~\cite{aAudit01} has revealed that LRC in the $10-200~bp$ range are the signature 
of the nucleosomal structure and that over larger distances ($\geq 200~bp$) 
they are likely to play a role in the condensation 
of the nucleosomal string into the $30~nm$-chromatin 
fiber. To which extent sequence-dependent LRC structural disorder does 
help to regulate the structure and dynamics of chromatin 
is of fundamental importance as regards to 
the potential structural informations that may have been 
encoded into DNA sequences during evolution. A possible key to the 
understanding is that the LRC structural disorder induced by the sequence 
may favor the formation of small (few hundreds $bp$) DNA 
loops and in turn the propensity of eukaryotic DNA to interact with 
histones to form nucleosomes.
\begin{figure}
\scalebox{0.7}{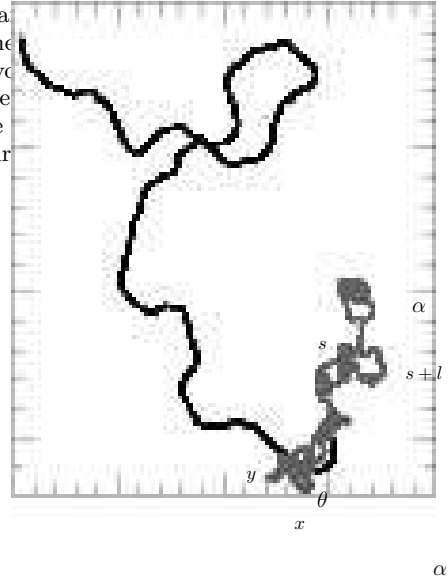}
\caption{Left: ``Spontaneous'' trajectory for two 2D semi-flexible chains with uncorrelated 
($H=0.5$, black) and LRC ($H=0.8$, grey) structural disorder. Right: ``winding'' constraint (top) and 
cyclization constraint (bottom).}
\label{fig_traj}
\end{figure}

Our aim here is to investigate the influence of LRC 
structural disorder on the thermodynamical properties of semi-flexible 
chains like DNA when constrained locally to form a loop of size $l$ much 
smaller than the chain length $L$. Because of the 
approximate planarity of nucleosomal DNA loops, one will assume the 
chains to be confined in a plane and to be free of 
any twisting deformation. Within the linear elasticity approximation, 
the local elastic energy variation of a 2D semi-flexible chain is: 
\begin{equation}
\delta E(s)=\tilde{A}(\dot{\theta}(s)-\dot{\theta_o}(s))^2/2,
\end{equation}
where $\tilde{A}$ is the bending stiffness,  $\dot{\theta}(s)$ the local curvature and $\dot{\theta_o}(s)$ 
the local ``spontaneous'' curvature of the chain. To model the intrinsic 
quenched ($T=0$) disorder, we consider 
$\dot{\theta_o}(s)$ as the realization of a gaussian fractional noise of 
zero mean and variance $\sigma_o^2$ and such that the corresponding random walk  
\normalsize$\Delta \Theta_o(s,l)=\int_s^{s+l} \dot{\theta}_o(u) du$\normalsize~exhibits normal fluctuations 
characterized by: 
\begin{equation}
\overline{\Delta \Theta_o}(s,l)= 0,~~\overline{\Delta \Theta^2_o}(s,l)-
\overline{\Delta \Theta_o}^2(s,l)= \sigma_o^2 l^{2H},
\label{eq_stat}
\end{equation}
where $H$ is called the Hurst exponent~\cite{aAudit01,aVaill03}: when $H=1/2$, one recovers the standard 
uncorrelated gaussian  noise, 
and for $H >1/2$, the distribution of the intrinsic curvature along the chain is LRC. 
As illustrated in Fig.~\ref{fig_traj}, due to the persistence 
of the orientation's fluctuations, LRC 2D spontaneous trajectories are 
more looped than the uncorrelated ones.

\begin{figure}
\scalebox{0.5}{\includegraphics{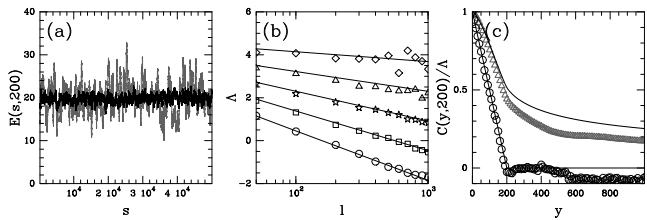}}
\caption{Free energy of a single loop of size $l$ 
in a chain of length $L=10^5$ under the ``winding'' constraint and for 
$A=200$, $\sigma_o=0.01$ and $\alpha= 2 \pi$. (a) Energy landscape 
along a chain with uncorrelated ($H=0.5$, black) and LRC ($H=0.8$, grey) 
disorders. (b) Free energy r.m.s. fluctuations $\Lambda(l)$ {\em vs.} $l$; 
the symbols correspond to numerical estimate 
for different disorders: $H=0.5$ ($\circ$), $0.6$ ($\square$), $0.7$ (\Pisymbol{pzd}{73}), 
$0.8$ ($\triangle$) and $0.9$ (\small$\Diamond$\normalsize); 
the solid lines correspond to Eq.~(\ref{fm}). 
(c) Reduced correlation function $C(y,200)/\Lambda(200)$ 
{\em vs.} $y$; the symbols have the same meaning as in (b); 
the solid curves correspond to Eq. (\ref{eq_corr}).}
\label{fig_stat}
\end{figure}
To account for the spontaneous formation of loop of size $l$, we will consider chains 
under the following geometrical constraints (Fig.~\ref{fig_traj}): (i) the 
``winding'' constraint which amounts to keep fixed the variation of the orientation 
over a length $l$, \small$\int_s^{s+l} \dot{\theta}(u) du =  \alpha$\normalsize, and (ii) 
the ``cyclization'' constraint where in addition 
the two extremities are held fixed together, \small $\int_s^{s+l} \cos(\theta(u)) du 
= \int_s^{s+l} \sin(\theta(u)) du = 0$\normalsize. Given a chain defined by its spontaneous curvature distribution, we first compute 
the 1D energy landscape $E(s,l)$ associated to the formation of one loop of length $l$ 
at the position $s$. Introducing the constraint via Lagrange multipliers, the 
equilibrium configuration is obtained by solving the corresponding Euler-Lagrange 
equations. For the ``winding'' constraint, from the equilibrium equations, one gets 
immediately the shape of the constrained chain and the corresponding energy cost 
\begin{equation}
E(s,l) =  \tilde{A}[\Delta\Theta^2_o(s,l) - 2 \alpha \Delta\Theta_o(s,l)
+\alpha^2]/2l.
\label{eq_enth}
\end{equation}
In Fig.~\ref{fig_stat}(a) are shown  the energy 
landscapes for $l=200$ of an uncorrelated and a LRC chains; the 
fluctuations of the later are of much larger amplitude than those of the former. 
In the weak disorder (WD) limit ($\sigma_o^2 \ll 1$), the statistics of the energy landscape 
is gaussian; when using Eq.~(\ref{eq_stat}), one gets for the 
mean \small$\overline{E}(l) = \tilde{A}[\alpha^2/l + \sigma^2_o l^{2H-1}]/2$\normalsize~and the 
variance \small$\overline{(E(l)-\overline{E}(l))^2}=\alpha^2 \tilde{A}^2 \sigma_o^2 l^{2H-2}$\normalsize. 
For the ``cyclization'' constraint there is no such general 
analytic derivation of the equilibrium configuration and one has to turn back to 
numerical computations. As in~\cite{aZhang03}, we have used an iterative scheme to 
perform numerical computations for several values of $\alpha,~H,~\sigma_o$ and 
$l$. In the WD limit, the equilibrium energy fluctuations 
numerically obtained with the ``cyclization'' constraint, display gaussian statistics 
with the same mean and variance as previously derived with the ``winding'' constraint. 

At finite temperature, one has to consider the effect of thermal fluctuations which requires 
to compute the free energy cost of the loop formation 
\small$\beta f(s,l)= \beta E(s,l)- \Delta S(s,l)$\normalsize, 
where \small$\beta=1/k_BT$\normalsize. Under harmonic approximation, the entropy cost, 
\small$\Delta S (l)= b - c \ln{l}$\normalsize, can be computed analytically (resp. numerically) 
for the ``winding'' (resp. ``cyclization'') constraint $c_w=1/2$ (resp. $c_c=7/2$). 
We finally get the following free energy landscape statistical properties in the WD limit:
\begin{eqnarray}
&&\beta\overline{f}(l) = \frac{A}{2}[\frac{\alpha^2}{l}+ \sigma^2_o l^{2H-1}] + c\ln{l} -b,  \label{fm}\\
&&\beta^2\overline{\left(f(l)-\overline{f}(l)\right)^2} = A^2 \alpha^2 \sigma_o^2 l^{2H-2} \overset{not.}{=} \Lambda(l). \notag
\end{eqnarray}
But the thermodynamical properties of the system are 
likely to depend on the correlations of the free energy landscape. From Eq.~(\ref{eq_enth}), one gets:
\begin{equation}
  \hspace{-0.1ex}
  C(s'-s,l)= \beta^2\overline{\delta f (s',l) \delta f(s,l)}=\Lambda (l)\mathcal{C}_{H}\left(\frac{s'-s}{l}\right),
\label{eq_corr}
\end{equation}
where \small$\delta f (s,l)=f(s,l)-\overline{f}(l)$\normalsize~and  
\small$\mathcal{C}_{H}(y)=(|y+1|^{2H}+|y-1|^{2H}-2|y|^{2H})/2$\normalsize~is the 
correlation function of fractional Brownian motions (fBm)~\cite{aMande77}. The results reported 
in Fig.~\ref{fig_stat}(b) show that the scaling form (\ref{fm}) of the free energy 
r.m.s. fluctuations is well verified for weak disorder ($\sigma_o=0.01$) up to loop 
size $l \lesssim 10^3$. As shown in Fig.~\ref{fig_stat}(c), the free energy 
correlation function decreases rather fast over a distance of the order $l$, and then 
much slowly at larger distances (larger $H$, slower the decrease) in good agreement with 
the asymptotic behavior \normalsize$\mathcal{C}_{H}(y)\sim H(2H-1) y^{2H-2}$\normalsize~for 
\normalsize$y \to \infty$\normalsize~(Eq.~(\ref{eq_corr})). While the free energy fluctuations are short range correlated 
for $H=1/2$, they display LRC for $H>1/2$.

\begin{figure}
\scalebox{0.6}{\includegraphics{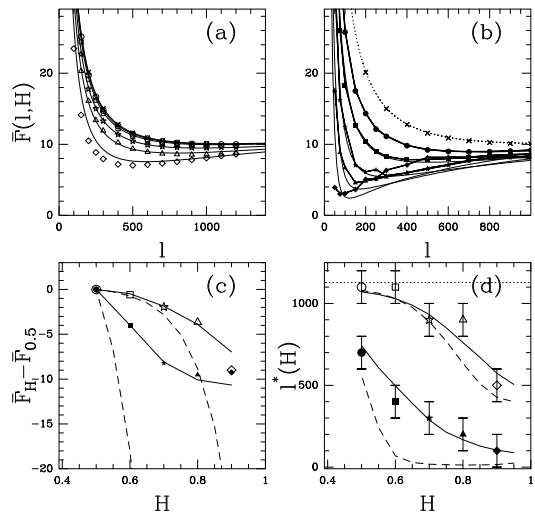}}
\caption{Free energy of the single loop system $\mathcal{F}_H(l,L)$ {\em vs} $l$ under the 
``cyclization'' constraint and for $L=15000,~A=200,~\alpha=2 \pi$ and $\sigma_o=0.01$ (a) 
and $0.05$ (b). The symbols correspond to a single 
chain's free energy $\mathcal{F}_H(l,L)$ obtained from the (exact) numerical computation of 
$f(s,l)$ for $H=0.5$ ($\circ,~\bullet$), $0.6$ (\small $\square,~\blacksquare$ \normalsize), 
$0.7$ (\small \Pisymbol{pzd}{73},\small \Pisymbol{pzd}{72}), 
$0.8$ ($\vartriangle,~\blacktriangle$) and $0.9$ (\small $\Diamond$, \ding{117}); 
the ($\times$) correspond to the ``pure'' case without disorder.
The continuous curves stand for the corresponding  quenched free energies $\overline{\mathcal{F}}_H(l,L)$ 
averaged over $100$ ``typical'' single chain free energies computed 
using Eq.~(\ref{eq_enth}) for the enthalpic part ({\em i.e.} the
``winding'' energy)  and $c_c= 7/2$ for the entropy 
cost; the dotted curve correspond to the exact analytical expression 
for the ``pure'' case. (c) $\mathcal{F}_{H}(l)-\mathcal{F}_{1/2}(l)$ {\em vs} $H$ between LRC 
and uncorrelated chains, for loop size $l=200$; 
the dashed curves correspond to perturbative approximation. 
(d) Optimal loop length $l^{\ast}(H)$ {\em vs} $H$; the dashed curves correspond to the perturbative 
expression (\ref{eq_lopt}); 
the horizontal line indicates the optimal loop length $l^*=1128$ for the ``pure'' system.
In (c) and (d) the symbols and the continuous curves have 
the same meaning as in (a) and (b)} 
\label{fig_Fl}
\end{figure}

The thermodynamics of a single ``loop'' of size $l$ embedded in a 
chain of length $L$ is described by the partition function 
$Z(l,L)= \int_{0}^{L-l} \exp{\left[-\beta f(s,l)\right]} ds$, which accounts for 
all the possible locations of the loop along the chain. The equilibrium properties 
are determined by the free energy of the system (relatively to the unconstrained state of the chain): 
$\beta \mathcal{F}(l,L)=-\ln{\left(Z(l,L)\right)}$. 
The thermodynamics associated to rugged energy landscapes have been widely studied during the 
past decades. The equilibrium and non equilibruum properties depend upon the statistics of 
energy fluctuations. When no correlations are present, it is the well known Random Energy Model (REM) that can be solved exactly~\cite{aDerri81}. 
This model presents a freezing phase transition 
separating a \emph{self-averaging} ``high temperature'' (HT) phase where the ``constraint'' can explore all 
the possible configurations (positions) and a  ``low temperature'' (LT) phase 
dominated by the few lowest energy minima where the ``constraint'' is likely to be localized~\cite{aBouch97}.
But we have seen in Eq.~(\ref{eq_corr}) that the loop free energy fluctuations are LRC which may question 
the pertinence of the REM. In the HT/WD limit, $\beta \delta f(s,l) \ll 1,
~\forall s$, one gets for finite $L$:\begin{eqnarray}
&&\beta \overline{\mathcal{F}}(l) \simeq \beta \overline{f}(l) - \ln{(L-l)} - 
\frac{\beta^2}{2}\overline{\delta^2f}(l) +\frac{1}{2} \mathbf{C}(l,L), \label{eq_Fl} \\
&&\beta^2 \overline{\left(\mathcal{F}(l)-\overline{\mathcal{F}}(l)\right)^2} = 
\mathbf{C}(l,L)=\frac{1}{L^2}\iint C(s-s',l)dsds'. \notag
\end{eqnarray}
The correlations control the sample-to-sample fluctuations. An 
explicit computation gives for both the ``winding'' and ``cyclization'' constraints: 
$\mathbf{C}(l,L) \sim  \Lambda(l) \left(L/l)\right)^{2H-2} \propto L^{2H-2}$. The 
correlations vanish independently of $l$, in the thermodynamic limit $L \to \infty$ leading 
to the asymptotic validity of the REM~\cite{remCor}.
Combining Eqs.~(\ref{fm}) and (\ref{eq_Fl}), one gets in the HT/WD phase:
\begin{equation}
\beta \overline{\mathcal{F}}(l) =  \frac{A\alpha^2}{2l}+\frac{A\sigma^2_o}{2}l^{2H-1}-
\frac{A^2\alpha^2}{2}\sigma_o^2 l^{2H-2}+c\ln{l}-b-\ln{L}.
\label{eq_FlStat}
\end{equation}
In Figs.~\ref{fig_Fl}(a,b)  are reported the evolution of the free energy of the single 
loop system {\em vs}.\ the size of the ``cyclization'' constraint for 
$L=15000,~A=200,\alpha=2 \pi$ and a disorder amplitude $\sigma_o=0.01$ ($0.05$) 
comparable to that obtained when using experimentally established structural tables~\cite{aVaill03}). 
The symbols correspond to exact numerical estimation of the free energy 
for five values of $H$ that amount to strengthen LRC while the continuous 
curves correspond to the corresponding quenched free energy $\overline{\mathcal{F}}_H(l,L)$ averaged over 
$100$ single loop chains. From both numerical and analytical results, one can extract 
the following main messages: (i) In the absence of disorder ($\sigma_o=0$), the ``pure'' system has 
a free energy that presents a minimum for a finite length $l^*=\alpha^2A/2c$ (Eq.~(\ref{eq_lopt})). 
This optimal length separates the enthalpic domain at small scale $l$ characterized by a power law decrease 
of the free energy, 
and the entropic domain at large scale characterized by a logarithmic increase. (ii) 
When one adds some intrinsic uncorrelated disorder ($H=1/2$), the $l$-dependence of the free energy 
reduces (up to a constant) to an homogeneous ``pure'' case with a renormalized value of the 
bending flexibility $A_{eff}=A(1-A\sigma_o^2)$~\cite{aVaill03,ibTrifo87}. 
Thus there is no qualitative difference between an uncorrelated system and an ideal one, but introducing disorder 
decreases the free energy (Fig.~\ref{fig_Fl}(a,b)) and favors the formation of 
loop of smaller size $l^*=\alpha^2 A_{eff} /2c$. 
(iii) When considering LRC disorder, then the  system no longer behaves as an homogeneous one, but 
more importantly, in the small scale domain, both the free energy and the optimal length $l^*(H)$ decrease 
(Fig.~\ref{fig_Fl}(d)) when one increases $H$. 
As shown in Fig.~\ref{fig_Fl}(c), for a fixed loop size 
$l=A=200$, the quenched average free energy provides a good description of the free energy 
of a typical single loop chain for both $\sigma_o=0.01$ and $0.05$. Note that only for 
$\sigma_o=0.01$ and value of $H \lesssim 0.7$, these results are well accounted by the HT/WD 
approximation (Eq.~(\ref{eq_FlStat})). 
Similar results are obtained in Fig.~\ref{fig_Fl}(d) for the optimal loop length $l^*(H)$ which is shown to 
decrease down to values about a few hundreds when increasing $H$ from $0.5$ to $0.9$. For 
$\sigma_o=0.01$, the solution of the HT/WT perturbative equation:
\begin{equation}
 (2H-1)A\sigma^2_o l^{2H} +2c l-(2H-2)\alpha^2 A^2 \sigma_o^2 l^{2H-1}=A\alpha^2,
\label{eq_lopt}
\end{equation}
provides a rather good description of the $H$-dependence of the loop 
size $l^*(H)$ of a typical single loop chain. The perturbative expression of the free energy 
(Eq.~(\ref{eq_FlStat})) breaks down when the energy fluctuations 
become too large: this is the freezing 
transition towards the low temperature/strong disorder phase where the replica approach needs to 
be used to get the correct quenched free energy~\cite{aBouch97}. The computation of the localized 
states is not the purpose of this letter since as shown in Fig.~\ref{fig_Fl}, for 
parameter values compatible with DNA characteristic properties, namely $A=200,~\alpha=2 \pi$ 
and $\sigma_o \sim 0.01$, the HT/WD approximation is likely to apply.

As emphasized in Ref.~\cite{aSluts03}, a convenient formalism to investigate 
diffusion process in the random 1D potential $E(s,l)$ 
of the single fixed length loop is that of 
mean first passage time (MFPT). The MFPT (as expressed in number 
of elementary steps) at the position $N$ (starting from $s=0$) is given by:
\begin{equation}
\tau (N,l) \simeq 2 \int_0^N ds \int_{s}^{N} ds' \exp[2 \beta (E(s)-E(s'))].
\label{eq_tm}
\end{equation}
\begin{figure}
\scalebox{0.5}{\includegraphics{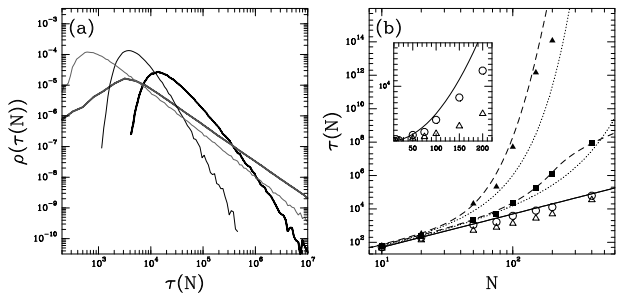}}
\caption{MFPT $\tau(N)$ (measured in number of steps) for $A=l=200,~\alpha=2 \pi,~L=15000$ and 
$\sigma_o=0.01$. (a) Pdf of $\tau(N)$ calculated for $100000$ uncorrelated ($H=0.5$, black) and 
LRC ($H=0.8$, grey) chains for $N=100$ (thin) and $200$ (thick).
(b) Most probable MFPT {\em vs} $N$ for $H=0.5$ ($\circ$) and $0.8$ ($\vartriangle$); mean 
MFPT {\em vs} $N$ for $H=0.6$ ($\blacksquare$) and $0.8$ ($\blacktriangle$); the dashed curves 
correspond to the analytical quenched average  (Eq.~(\ref{eq_tmanal})); 
the dotted curves correspond to the 
small displacement approximation (Eq.~(\ref{eq_tm_petit})); 
the continuous line correspond to the ``pure'' diffusion case $\tau(N)=N^2$.} 
\label{fig_tmean}
\end{figure}
The average over all possible realizations of the disordered energy landscape leads to:\begin{equation}
\overline{\tau} (N,l) \simeq 2 \int_0^N ds \int_{s}^{N} ds'e^{4 \Lambda(l) \left(1 - \mathcal{C}_H(\frac{s'-s}{l})\right)}.
\label{eq_tmanal}
\end{equation} 
When looking at displacements smaller or of the order of the loop size, $N \lesssim l$, 
then the typical 
energy barrier increases like $\Delta E(N) \sim N^H$: the energy landscape has a fBm structure. 
For $(s'-s)/l \ll 1$, Eq.~(\ref{eq_tmanal}) reduces to:
\begin{equation}
\overline{\tau}(N,l)\sim N^2 e^{\frac{4}{(2H+1)(H+1)} \Lambda(l)(N/l)^{2H}}.
\label{eq_tm_petit}
\end{equation} 
We thus get  a stretched exponential creep that depends on $H$. For $H=1/2$, one recovers the exponential 
creep of the Random Force Model (RFM) with logarithmically slow (``Sina\"i'') diffusion~\cite{aBouch90,aHwa03}. 
When strengthening the LRC by increasing $H >1/2$, one further increases 
$\overline{\tau}(N,l)$ suggesting some slowing down of the loop dynamics. In 
Fig.~\ref{fig_tmean}(b), this modified Sina\"i diffusion~\cite{aBouch90} 
accounts quite well for the short distance dynamics of single 
loop chain realizations. But as shown in Fig.~\ref{fig_tmean}(a), when computing 
the probability density function (pdf) of the MFPT over 
$100000$ realizations for distances $N \lesssim 2l$, the way the average 
MFPT depends on $H$ is very much affected by the evolution of the pdf tail and 
does not reflect the dependence of the most probable MFPT (as defined by 
the pdf maximum) which in contrast decreases when increasing $H$. This shows that for a typical 
event, the motion of a single loop in a LRC chain over distances of the 
order of its size is definitely superdiffusive (see inset Fig.~\ref{fig_tmean}(b)), 
larger $H$, faster the dynamics.

To summarize, we have shown that the competing effects of entropy and sequence dependent 
structural disorder favors the autonomous formation of DNA loops. When taking into 
account the existence of LRC as observed in eukaryotic genomic sequences~\cite{aAudit01}, 
we have found, in the WD limit, that strengthening LRC allows the formation of smaller 
loops that superdiffuse, larger the LRC, faster the typical local loop dynamics. 
These results strongly suggest that these LRC may have been encoded into genomic sequences 
during evolution to predispose eukaryotic DNA to interact with histones to form nucleosomes. 
The size of the selected loops (few hundreds $bp$) are typical of the characteristic DNA which is wrapped around 
histones; we refer the reader to a recent work of Bussiek {\em et al.}~\cite{aBussi05} 
where in high salt concentration conditions, the nucleosomes are observed to be preferentially 
located at the crossing of DNA loops of characteristic length $\sim 200~bp$ ($50~nm$). The local 
rapid diffusion of the loop induced by the LRC structural disorder provides 
a very attractive interpretation to the nucleosome 
repositioning dynamics. LRC are likely to help the nucleosomes to rearrange themselves in a very 
efficient way as, {\em e.g.} after the passage of the 
transcription and replication polymerases. Since in {\em in vivo} chromatin, the 
nucleosomal string presents a high occupation density with an average distance between nucleosomes of the order of 
$50~bp$, this raises the issue of the effect of the interaction between nucleosomes on their 
large scale mobility. The generalization of the present work to multiple 2D loops in a long 
LRC DNA chain is in current progress.

\bibliographystyle{prsty}

\begin{thebibliography}{10}
\bibitem{remCor}
Note that 
for $H \to 1/2$, one recovers the $1/L$ convergence to zero of the 
correlation function obtained for un- (and short-range) correlated free energy landscapes.
\end{thebibliography}

\end{document}